\documentclass[10pt, conference]{IEEEtran}
\IEEEoverridecommandlockouts
\usepackage{cite}
\usepackage{amsmath,amssymb,amsfonts}
\usepackage{algorithmic}
\usepackage{graphicx}
\usepackage{textcomp}
\usepackage{siunitx}
\usepackage{xcolor}
\usepackage{listings}
\usepackage{tabularx}
\usepackage{booktabs}
\usepackage{csquotes}
\usepackage[most]{tcolorbox}
\usepackage{multirow}
\usepackage{hyperref}
\usepackage{orcidlink}
\usepackage{balance}

\usepackage{xurl}
\usepackage{cleveref}
\usepackage{enumitem}
\usepackage{subcaption}
\usepackage{multirow}
\usepackage{adjustbox}
\urldef\replicationA\url{https://github.com/seer-lab/FlakyXbert/blob/main/src/IDoFT%20dataset%20code/FlakyXbert-IDoFT_binary_projectwise_nifi.ipynb}
\urldef\replicationB\url{https://github.com/seer-lab/FlakyXbert/blob/main/src/IDoFT%20dataset%20code/FlakyXbert-IDoFT_binary_projectwise_hadoop.ipynb}
\urldef\replicationC\url{https://github.com/seer-lab/FlakyXbert/blob/main/src/IDoFT%20dataset%20code/FlakyXbert-IDoFT_binary_projectwise_fastjson.ipynb}
\urldef\replicationD\url{https://github.com/seer-lab/FlakyXbert/blob/main/src/IDoFT%20dataset%20code/FlakyXbert-IDoFT_binary_projectwise_admiral.ipynb}
\urldef\replicationE\url{https://github.com/seer-lab/FlakyXbert/blob/main/src/IDoFT%20dataset%20code/calculate.ipynb}

\newcommand\Cshadowbox{\VerbBox\@Cshadowbox}
\def\@Cshadowbox#1{%
  \setbox\@fancybox\hbox{\fbox{#1}}%
  \leavevmode\vbox{%
    \offinterlineskip
    \dimen@=\shadowsize
    \advance\dimen@ .5\fboxrule
    \hbox{\copy\@fancybox\kern.5\fboxrule\lower\shadowsize\hbox{%
      \color{ShadowColor}\vrule \@height\ht\@fancybox \@depth\dp\@fancybox \@width\dimen@}}%
    \vskip\dimexpr-\dimen@+0.5\fboxrule\relax
    \moveright\shadowsize\vbox{%
      \color{ShadowColor}\hrule \@width\wd\@fancybox \@height\dimen@}}}
\makeatother
\colorlet{ShadowColor}{gray}

\lstdefinelanguage{javadiff}{
  language=Java,
  morecomment=[f][\color{red}]{-},
  morecomment=[f][\color{green!50!black}]{+},
}

\lstset{
  basicstyle=\ttfamily\small,
  showstringspaces=false,
}

\lstset{
  basicstyle=\ttfamily\small,  
  frame=single,                
  numbers=none,                
  captionpos=b,                
  xleftmargin=10pt,            
  xrightmargin=10pt            
}

\newtcolorbox{promptbox}[2][]{
  colback=gray!5!white,
  colframe=gray!75!black,
  fonttitle=\bfseries, 
  title={#2}, 
  #1
}

\def\BibTeX{{\rm B\kern-.05em{\sc i\kern-.025em b}\kern-.08em
    T\kern-.1667em\lower.7ex\hbox{E}\kern-.125emX}}
\begin{document}

\title{Can We Classify Flaky Tests Using Only Test~Code? An LLM-Based Empirical Study}

\author{\IEEEauthorblockN{Alexander Berndt \orcidlink{0009-0009-5248-6405}$^{*}$ }
\IEEEauthorblockA{
 \textit{Heidelberg University} \\
alexander.berndt@uni-heidelberg.de
} \\
\IEEEauthorblockN{Marcus Kessel \orcidlink{0000-0003-3088-2166}}
\IEEEauthorblockA{\textit{University of Mannheim} \\
marcus.kessel@uni-mannheim.de
}\\
\and
\IEEEauthorblockN{Vekil Bekmyradov \orcidlink{0009-0007-7429-8259}}
\IEEEauthorblockA{ \textit{SAP} \\
vekil.bekmyradov@sap.com
} \\
\IEEEauthorblockN{Thomas Bach \orcidlink{0000-0002-9993-2814}}
\IEEEauthorblockA{\textit{SAP} \\
thomas.bach03@sap.com 
}
\and
\IEEEauthorblockN{Rainer Gemulla \orcidlink{0000-0003-2762-0050}}
\IEEEauthorblockA{\textit{University of Mannheim} \\
rgemulla@uni-mannheim.de
} \\
\IEEEauthorblockN{Sebastian Baltes \orcidlink{0000-0002-2442-7522}}
\IEEEauthorblockA{\textit{Heidelberg University} \\
sebastian.baltes@uni-heidelberg.de 
}
\thanks{$^*$Also affiliated with SAP.}
}

\maketitle

\begin{abstract}
Flaky tests yield inconsistent results when they are repeatedly executed on the same code revision. 
They interfere with automated quality assurance of code changes and hinder efficient software testing.
Previous work evaluated approaches to train machine learning models to classify flaky tests based on identifiers in the test code. 
However, the resulting classifiers have been shown to lack generalizability, hindering their applicability in practical environments.
Recently, pre-trained Large Language Models (LLMs) have shown the capability to generalize across various tasks.
Thus, they represent a promising approach to address the generalizability problem of previous approaches.

In this study, we evaluated three LLMs (two general-purpose models, one code-specific model) using three prompting techniques on two benchmark datasets from prior studies on flaky test classification. Furthermore, we manually investigated 50 samples from the given datasets to determine whether classifying flaky tests based only on test code is feasible for humans.

Our findings indicate that LLMs struggle to classify flaky tests given only the test code. The results of our best prompt-model combination were only marginally better than random guessing. In our manual analysis, we found that the test code does not necessarily contain sufficient information for a flakiness classification.
Our findings motivate future work to evaluate LLMs for flakiness classification with additional context, for example, using retrieval-augmented generation or agentic AI. 
\end{abstract}


\section{Introduction}
\label{ch:flaky-test-detection}

Flaky tests yield inconsistent results when they are repeatedly
executed on the same code revision~\cite{parry2021survey}. They interfere with continuous integration pipelines, hamper efficient automated software testing, and diminish developers’ trust in the reliability of test results~\cite{berndt2024test}. Previous work has proposed various approaches to detect and fix flaky tests~\cite{eck2019understanding,luo2014empirical,parry2021survey,fatima2024flakyfix,liu2024WEFix,berndt2023vocabulary,berndt2024test,magill2025deflake,haben2024importance}. Typically, flaky tests can be detected by repeatedly executing a test and analyzing its results. However, repeatedly executing tests is costly with respect to computational resources. 

A common approach to detect flaky tests without repeated test execution is to train a machine learning model on the task of flakiness classification~\cite{hoang2024presubmit,haben2024importance,lampel2021life,more2025analysis,pinto2020vocabulary,verdecchia2021know,haben2021replication,camara2021vocabulary,berndt2023vocabulary,rahman2024quantizing,fatima2022flakify,eck2019understanding}. For example, Pinto et al. proposed training machine learning classifiers on bag-of-words representations of the test code with promising
results~\cite{pinto2020vocabulary}. Additional studies evaluated various models, training approaches, and evaluation setups for the task of flakiness classification based on test code~\cite{fatima2022flakify,berndt2023vocabulary,camara2021vocabulary,haben2021replication,rahman2024quantizing}. Although the resulting classifiers generally achieved strong results on established benchmarks, evaluations conducted under conditions resembling real-world usage indicated limited generalizability caused by an overfitting of the classifiers to the test code vocabulary in the training set~\cite{berndt2023vocabulary,haben2021replication,camara2021vocabulary}. 

To gain a deeper understanding of flakiness classification based on the test code, in this study, we investigated the following research questions: 

\textbf{RQ1:} What is the performance of LLMs without additional fine-tuning in flaky test classification based on the test code...

\phantom{    }\textbf{(1.1)} ...with a zero-shot prompt?

\phantom{    }\textbf{(1.2)} ...with a zero-shot chain-of-thought prompt?    

\phantom{    }\textbf{(1.3)} ...with a few-shot chain-of-thought prompt?

\textbf{RQ2:} What degree of non-determinism do we observe for LLMs when classifying flaky tests?

\textbf{RQ3:} To what degree do humans consider themselves capable of classifying flaky tests based on the test code?

Our results suggest that LLMs struggle to classify flaky tests solely on the basis of the test code. Furthermore, even with greedy decoding at temperature 0, the degree of non-determinism across repeated executions of our experiments complicates the deployment of such models in real-world scenarios for flakiness detection. Based on manual assessment, we conclude that test code alone is insufficient to detect certain types of flakiness. While there are issue types, such as the use of unordered collections, that may be apparent in the test code, we find that even humans require additional context information to identify more sophisticated flakiness issues. For example, flakiness may be caused by side effects of test utility functions whose functionality is not visible in the test code. 

In summary, our work provides the following contributions:
\begin{enumerate}
    \item Evaluation of three LLMs for flakiness classification on two common benchmarks.
    \item Discussion of implications for future work on flakiness~classification.
\end{enumerate}

The remainder of this document is structured as follows. 
\Cref{sec:datasets} contains relevant information on the datasets used.
We introduce relevant work in \Cref{sec:related} before we pinpoint existing issues in one prior study in \Cref{sec:prior}. We describe our methodology in \Cref{sec:questions-detection} and present our results in \Cref{sec:results-classification}. We discuss the results in \Cref{sec:discussion} and threats to validity in \Cref{sec:threats}, before concluding the paper in \Cref{sec:conclusions}. Our code and the experimental results are available online as part of our supplementary material~\cite{zenodo}.

\section{Datasets}
\label{sec:datasets}
We used two benchmark datasets for flakiness classification to evaluate LLMs for flakiness classification: the \emph{International Dataset of Flaky Tests} (IDoFT) and \emph{FlakeBench}~\cite{InternationalDatasetofFlakyTests,rahman2025understanding}.

\textbf{1.~IDoFT}:
The IDoFT dataset arises from an open-source GitHub repository where contributors track flaky tests in open-source Java and Python projects~\cite{InternationalDatasetofFlakyTests,LamETAL19iDFlakies}.
In this study, we utilized a subset of IDoFT that was previously used as a benchmark for binary flakiness classification~\cite {more2025analysis,rahman2024quantizing}.
This subset contains \num{3813} samples, of which 587 (15\%) were labeled non-flaky and \num{3226} (85\%) were labeled flaky.
Non-flaky samples represent flaky tests that have been fixed in a pull request with the status \enquote{Accepted}.
The samples in the dataset originate from 299 projects.
In addition to a label indicating whether a test is flaky, Fatima et al.~\cite{fatima2024flakyfix} have added category labels for the type of fix for each of the fixed tests. 
Labeling was performed using heuristics that Fatima et al. created based on a manual inspection of 100 samples.
\Cref{tab:fix-labels} provides an overview of the resulting fix labels and their prevalence in the given data.

\begin{table*}
\caption{Summary of the fix categories for flaky tests as identified by Fatima et al.~\cite{fatima2024flakyfix}. 
  Note that one test may have multiple fix labels. Therefore, the sum of the support does not match the sample count of the dataset.}
  \centering
  \begin{tabularx}{\textwidth}{l r X}
        \toprule
        Fix type & Support & Description  \\
        \midrule
        Change assertion & 175 &  Replacement of the employed assert function, e.g., \texttt{assertJsonStringEquals} instead of \texttt{assertEquals} \\
        Change condition & 109 & Replacement of a condition within the assert statement, e.g., replacing \texttt{containsExactly} with \texttt{containsExactlyInAnyOrder} \\
        Reset variable & 98 & Addition of a statement to reset the current state that was altered by the test, e.g., by calling a \texttt{clear} function \\
        Change data structure & 77 & Replacement of the employed data structure, e.g. \texttt{LinkedHashMap} instead of \texttt{HashMap} \\
        Reorder data & 39 & Addition of some sorting statement, e.g., calling \texttt{sort} on an array \\
        Miscellaneous & 37 & Everything else \\
        Reorder parameters & 35 & Change of parameter order in a function call \\
        Change data format & 30 & Change of the format of some variable, e.g., from string to object \\
        Call static method & 9 & Addition of a call to static method that encapsulates complex behavior, e.g., for test setup \\
        String matching & 5 & Change matching of string-encoded representations, e.g., for comparing JSON objects with a string \\
        Change time zone & 3 & Setting a specific time zone to avoid timing issues \\
        Sleep & 3 & Addition of a call to the \texttt{sleep} function, e.g., to avoid race conditions \\
        Handle timeout & 2 & Addition of a timeout value for proper timeout handling \\
        \bottomrule
  \end{tabularx}
  \label{tab:fix-labels}
\end{table*}



\textbf{2.~FlakeBench}: The FlakeBench dataset originates from 97 open-source repositories written in Java on GitHub. Rahman et al.~\cite{rahman2025understanding} repeatedly executed tests in the respective repositories 100 times to identify flaky tests. Given the results from 100 repeated executions, they labeled a test flaky if it yielded at least one passing and one failing result. This labeling approach resulted in a dataset consisting of \num{8574} tests, of which 280 were labeled flaky and \num{8294} non-flaky. 
\section{Related Work}
\label{sec:related}
Prior work proposed various approaches to detect flaky tests based on different datasets~\cite{hoang2024presubmit,haben2024importance,lampel2021life,herzig2015empirically,more2025analysis,pinto2020vocabulary,verdecchia2021know,haben2021replication,camara2021vocabulary,berndt2023vocabulary,rahman2024quantizing,fatima2022flakify}.
In this paper, we focus on related work that used the test code to predict whether a test is flaky~\cite{more2025analysis,pinto2020vocabulary,verdecchia2021know,haben2021replication,camara2021vocabulary,berndt2023vocabulary,fatima2022flakify,rahman2025understanding, rahman2024quantizing}.

Pinto et al.~\cite{pinto2020vocabulary} introduced the idea of using the test code for flakiness prediction based on the assumption that the test code of flaky tests exhibits syntactical patterns differentiating them from non-flaky tests. For example, flaky tests may be more likely to use certain words, such as \enquote{random} in identifier names. To test this assumption, they trained a random forest classifier on a bag-of-words representation of the test code. They evaluated their approach using the Matthews Correlation Coefficient (MCC), which ranges from -1 to 1, with 0 representing the average performance of random guessing and 1 representing a perfect classifier. With their approach, Pinto et al. achieved an MCC of 0.9 on the \emph{DeFlaker} benchmark~\cite{bell2018deflaker}, indicating a high correlation between the predicted labels and the ground truth. 

The original approach of Pinto et al. was replicated and evaluated multiple times~\cite{berndt2023vocabulary,camara2021vocabulary,haben2021replication}. Haben et al. evaluated the original approach on the same dataset with a different evaluation setup. They evaluated the original approach with a time-sensitive setup, in which the model was trained on code from older revisions and tested on newer revisions of the source code, resulting in a decrease in MCC of up to 21\%. Camara et al. also investigated the performance of the approach of Pinto et al.~\cite{pinto2020vocabulary} in a different evaluation setup, where the training data originated from software projects other than the test data. They observed a notable decrease in performance, achieving a prediction accuracy of only 0.48. Berndt et al.~\cite{berndt2023vocabulary} evaluated Pinto et al.'s approach in an industrial context. They found that a random forest classifier achieved an F1 score of 0.95 on an internal flakiness benchmark, which was derived from dedicated test executions for flakiness investigations, but failed to generalize to data from the production system. They concluded that the lack of generalizability prevented the classifier from being deployed in practice.

More recent work has evaluated fine-tuning LLMs for flakiness detection~\cite{more2025analysis,fatima2022flakify,rahman2024quantizing}. Fatima et al. utilized a fine-tuned version of \emph{CodeBERT}, a pre-trained language model with 125 million parameters, achieving an F1 score of  98\% on the IDoFT dataset~\cite{fatima2022flakify}. However, as pointed out in a follow-up study by Rahman et al.~\cite{rahman2025understanding}, their implementation suffered from data leakage, resulting in distorted results. The results without data leakage remain unknown. Rahman et al. applied quantization to the fine-tuned CodeBERT model resulting from Fatima et al.'s study to increase the efficiency during inference. They achieved an F1 score of up to 94\% on IDoFT while reducing training time and RAM usage by 25\% and 48\%, respectively. More et al.~\cite{more2025analysis} evaluated a few-shot learning approach in an intra-project scenario, where a model is trained and tested with data from a single software project. Thus, their approach reduced training time to 10\% while maintaining similar classification performance on the IDoFT dataset.

In contrast to previous work on flaky test detection, we explore the potential of LLMs without any task-specific fine-tuning to classify tests as flaky. Unlike previous work, our approach does not require task-specific training data but relies on the capability of recent LLMs to generalize across tasks. We conjecture that LLMs without additional fine-tuning could be a solution to overcoming the generalization issues that were previously mentioned~\cite{camara2021vocabulary,haben2021replication}.
\section{Issues of Prior LLM-based Flakiness Classifiers}
\label{sec:prior}
As described in \Cref{sec:related}, replications of previous work have already found that flakiness classifiers trained only on the test code do not generalize to new data from other projects~\cite{camara2021vocabulary} or future iterations of the same project~\cite{berndt2023vocabulary,haben2021replication}. For example, in a study by Berndt et al., the authors noted that the lack of generalization was caused by an over-reliance on spurious features and a lack of diversity in the given datasets~\cite{berndt2023vocabulary}. These existing replication studies used the original approach of Pinto et al.~\cite{pinto2020vocabulary}, which utilized a random forest classifier and a bag-of-words representation of the test code. However, we assume that newer studies using LLMs also suffer from spurious features and may therefore also struggle to generalize to new contexts.

To support this assumption, we analyzed the replication package of a recent study on flakiness classification by More et al.~\cite{replicationZ}.
Similar to other studies on LLM-based flakiness classification, More et al. used the IDoFT dataset to benchmark their approach~\cite{rahman2024quantizing,fatima2022flakify}.
They trained and evaluated their model in a per-project scenario, where a separate model was trained on 80\% of the data for each project and tested on the remaining 20\%.

A notable fraction of the projects in the IDoFT dataset is highly skewed towards flaky samples.
For example, in the notebook for the \texttt{nifi} project, we found that the resulting classifier achieved an F1 score of 91.5\% even though it classifies all tests as flaky~\cite{replicationA}.
In fact, we found that their approach yielded weighted F1-scores greater than 90\% in 4 of 12 evaluated projects without correctly classifying any non-flaky samples~\cite{replicationB,replicationB,replicationC,replicationD}. Therefore, we conclude that the classifier did not learn meaningful patterns to classify tests.

Based on an analysis of the remaining projects, we observed that the samples in a per-project scenario show low diversity. 
For example, for the two remaining projects with the highest support, \texttt{junit-quickcheck} and \texttt{spring-data-r2dbc}, almost all non-flaky samples shared the same piece of code fixing the flakiness.
\Cref{sub:a} illustrates an example of the \texttt{junit-quickcheck} project, which had the highest support in the dataset. 
In this case, the addition of the line \texttt{Enums.iterations=0;} distinguishes the non-flaky version of the test from the flaky version.
We observed that this statement holds for the majority of samples in this project. Thus, classifying only based on the line \texttt{iterations = 0} yields an F1 score of 96\% when evaluated in the per-project scenario used in the study.

Therefore, we conclude that their approach is likely to suffer from the same problems pointed out in previous studies on the weaknesses of flakiness classifiers based on the vocabulary of the test code. More and Jeremy reported a weighted F1 score of 95.1\% across the projects in the IDoFT dataset~\cite{more2025analysis}. However, based on the result in a Jupyter notebook in the replication package~\cite{replicationE}, we found that the actual weighted F1 score is 88.1\%. Upon notification, the authors acknowledged this mistake in the paper and confirmed the reported F1 score in the notebooks.

\section{Methodology}
\label{sec:questions-detection} 
In this section, we describe our methodology for answering the research questions.

\begin{table}
  \caption{Landis et al.'s interpretation of Cohen's Kappa~\cite{landis1977measurement}.}
  \centering
  \begin{tabular}{r l}
    \toprule
    Values & Interpretation \\
    \midrule
    $\leq 0$ & Poor \\
    $[0.01,0.2]$ & Slight \\
    $[0.21, 0.4]$ & Fair \\
    $[0.41, 0.6]$ & Moderate \\
    $[0.61, 0.8]$ & Substantial \\
    $[0.81, 1]$ & Almost perfect \\
    \bottomrule    
  \end{tabular}
  \label{tab:cohens}
\end{table}

\subsection{RQ1: Classification}
 For the first research question, we used three models, \texttt{GPT-4o}, \texttt{GPT-OSS-120b}, and \texttt{Qwen3-Coder-480b}~\cite{hurst2024gpt,openai2025gptoss120bgptoss20bmodel,qwen3technicalreport}. We used all models with a temperature of 0 throughout the study to minimize the inherent non-determinism of the models~\cite{ouyang2025empirical}. We deployed \texttt{GPT-OSS-120b} and \texttt{Qwen3-Coder-480b} on a machine with 8 NVIDIA H200 GPUs using vLLM~\cite{kwon2023efficient}. We evaluated four different settings for the prompt, as prompt engineering has been shown to improve the performance of LLMs for various tasks~\cite{brown2020language,kojima2022large}. We differentiated the three settings for our prompt based on the components that constitute it. \Cref{fig:prompts} illustrates our prompt template.

\textbf{Task description}: 
The task description $x_{desc}$ describes the task to be performed by the LLM.
For our case, we used the task \emph{Classify the test as either flaky or not}.

\textbf{Instructions for implicit reasoning}:
Based on OpenAI's prompt engineering guidelines~\cite{openai-guidelines}, we added a set of instructions containing intermediate reasoning steps $x_{reason}$ to approach the problem of binary flakiness classification. More specifically, we instructed the model to reflect on potential issues in the test code before providing the answer.

\textbf{Demonstrations}:
Previous work has demonstrated that LLMs can learn from demonstrations within a given context through in-context learning~\cite{lampinen2022can}.
That is, given a set of labeled samples in the context, an LLM can infer the underlying pattern and apply it to new inputs without any parameter updates.
To benefit from this capability, we added $k=6$ annotated samples of test code $t$ to the prompt. 
We added three flaky ($y=1$) and three non-flaky ($y=0$) samples to avoid biasing the model towards one of the two classes, such that $x_{demo}=\{(t_1, y_{t_1}), ...,(t_6, y_{t_6})\}$. 
For each of the demonstrations in $x_{demo}$, we added the following three intermediate reasoning steps explaining the thought process to reach a conclusion on $y$ given $x_t$:
\begin{itemize}[label=--]
  \item A semantic description of the test code.
  \item An elaboration on the critical lines related to flakiness.
  \item The conclusion on whether the test is considered flaky.
\end{itemize}
Based on these three components, we defined the three settings for our prompt that we evaluated in this study as follows:
\begin{enumerate}
  \item $x_{zero} = \{x_{desc}\}$
  \item $x_{zero-CoT} = \{x_{desc}; x_{reason}\}$
  \item $x_{few-CoT} = \{x_{desc}; x_{reason}; x_{demo}\}$
\end{enumerate}
\Cref{fig:prompts} illustrates the structure of $x_{CoT}$.
\begin{figure}
    \centering  
    \begin{promptbox}{Exemplified $x_{CoT}$}
    Role: Expert Software Engineer.
    
    Task: Classify the provided test as flaky or not.
    
    Instructions:
    \begin{enumerate}[topsep=0pt]
      \itemsep0em
        \item Analyze the \verb|<CODE>| segment, which contains an existing test.
        \item Think about ways to improve the test or to fix existing issues with the test code.
        \item Follow the reasoning style shown in the examples to identify potential sources of flakiness. Think step by step.
        \item Classify the test as either flaky or not based on your thoughts on existing issues.
        \item In your response, only include the following labels, refrain from using any natural language.
            \begin{itemize} 
                \item[] 0: the test is not flaky
                \item[] 1: the test is flaky
            \end{itemize}
    \end{enumerate}
    
    Examples: \newline
    \-\hspace{.5cm}\textless EXAMPLE\_CODE\textgreater \newline
    \-\hspace{1cm}\textcolor{purple}{\{example\_code\}} \newline
    \-\hspace{.5cm}\textless /EXAMPLE\_CODE\textgreater
    
    \-\hspace{.5cm}\textless THOUGHTS\textgreater
    \begin{itemize}[topsep=0pt,label=--]
        \itemsep0em
        \addtolength{\itemindent}{.5cm}
        \item \textcolor{purple}{\{semantic\_test\_description\}}
        \item \textcolor{purple}{\{flakiness\_description\}}
        \item \textcolor{purple}{\{conclusion\}}
    \end{itemize}
    \-\hspace{.5cm}\textless /THOUGHTS\textgreater
    
    \-\hspace{.5cm}\textless ANSWER\textgreater \newline
    \-\hspace{1cm}\textcolor{purple}{\{label\}} \newline
    \-\hspace{.5cm}\textless /ANSWER\textgreater \newline \newline
    [... additional examples]
    \newline
    \newline
    \textless CODE\textgreater \newline
    \-\hspace{.5cm}\textcolor{purple}{\{code\}} \newline
    \textless /CODE\textgreater
    \end{promptbox}
    \caption{The structure of our $x_{CoT}$ prompt template.}
    \label{fig:prompts}
\end{figure}

To compare our results with those of previous studies, we calculated the weighted precision, weighted recall, and the weighted F1 score as metrics~\cite{more2025analysis}.
As a baseline, we also calculated the results for classifying all tests as flaky and a random prediction with 50\% chance of classifying a test as flaky.
Furthermore, we added MCC as an evaluation metric to account for the skewed class distributions in the datasets.

\subsection{RQ2: Non-Determinism}
We compared two result vectors from an LLM obtained in two repeated experiments using greedy decoding with a temperature of 0. 
For example, classifying $n$ tests as flaky (or not) results in a vector of length $n$, containing only 0 (non-flaky) and 1 (flaky). 
We measured the distance between two of such vectors using the normalized Hamming distance~\cite{sun2025emperor}.
The Hamming distance $H$ measures the distance between two vectors $x$ and $y$ of length $n$ as follows:
\begin{equation}
  H(x, y) = \frac{1}{n} \sum_{i=1}^n 1 [x_i \neq y_i]
\end{equation}
That is, intuitively, the Hamming distance measures the number of transitions from 0 to 1 required to transform one vector into the other.

\subsection{RQ3: Human Judgement}
In our third research question, we aimed to determine whether humans are capable of classifying tests as flaky solely based on the test code.
To achieve this, we conducted a manual survey of $n=50$ flaky examples from the IDoFT dataset, sampled from the subset provided by Fatima et al.~\cite{fatima2024flakyfix} as described in \Cref{sec:datasets}.
Based on the question \enquote{How likely would a developer proficient in Java classify this test as flaky?}, we labeled tests on a five-point Likert scale item ranging from 1 (Very likely) to 5 (Very unlikely).
Furthermore, we included an open text field for the annotators to provide additional comments.

To create a common understanding between the two reviewers, we preceded the labeling with an alignment step.
That is, we selected a preliminary set of 10 random samples and discussed the methodology to be used for labeling.
Thus, the reviewers agreed on the following framework:
\begin{itemize}
  \item The time box is five minutes per sample.
  \item Reviewers may look up only Java or test framework-specific knowledge if necessary (i.e., no project-specific knowledge).
  \item If the problem in the test code is not obvious at first sight, reviewers may look up the existing fix before labeling.
\end{itemize} 

We quantified the inter-rater agreement between the two reviewers using Cohen's Kappa, as implemented by \texttt{scikit-learn}~\cite{cohen1960coefficient}.
The values of Cohen's Kappa are within the interval $[-1, 1]$. \Cref{tab:cohens} shows the interpretation of Cohen's Kappa values.
We applied the weighted version of Cohen's Kappa, using quadratic weights, to account for the ordinal nature of the rating scale.

\begin{table}
    \caption{Flakiness classification results on the IDoFT dataset for each prompt, separated by model. The numbers in bold represent the highest value in a column.}
    \centering
    \begin{tabular}{l l r rrrr}
      \toprule
      \textbf{Approach} & \textbf{Model} & \textbf{Iter.} & \textbf{Prec.} & \textbf{Rec.} & \textbf{F1} & \textbf{MCC} \\
      \midrule
      
      \multirow{9}{*}{$x_{zero}$}
      & \multirow{3}{*}{GPT-4o} 
          & 1 & 0.88 & 0.14 & 0.24 & 0.03 \\
      &   & 2 & 0.87 & 0.14 & 0.24 & 0.03 \\
      &   & 3 & 0.88 & 0.14 & 0.24 & 0.04 \\
      \cmidrule(lr){2-7}
      & \multirow{3}{*}{GPT-OSS}
          & 1 & 0.90 & 0.29 & 0.44 & 0.10 \\
      &   & 2 & 0.91 & 0.30 & 0.43 & 0.12 \\
      &   & 3 & 0.91 & 0.30 & 0.45 & 0.12 \\
      \cmidrule(lr){2-7}
      & \multirow{3}{*}{Qwen-Code}
          & 1 & 0.91 & 0.26 & 0.41 & 0.11 \\
      &   & 2 & 0.91 & 0.26 & 0.41 & 0.10 \\
      &   & 3 & 0.91 & 0.26 & 0.41 & 0.11 \\
      \midrule
      \multirow{9}{*}{$x_{zero-CoT}$}
      & \multirow{3}{*}{GPT-4o}
          & 1 & 0.84 & 0.78 & 0.81 & 0.00 \\
      &   & 2 & 0.84 & 0.79 & 0.81 & -0.03 \\
      &   & 3 & 0.85 & 0.79 & 0.82 & 0.01 \\
      \cmidrule(lr){2-7}
      & \multirow{3}{*}{GPT-OSS}
          & 1 & 0.87 & 0.58 & 0.70 & 0.06 \\
      &   & 2 & 0.86 & 0.58 & 0.69 & 0.05 \\
      &   & 3 & 0.86 & 0.56 & 0.68 & 0.05 \\
      \cmidrule(lr){2-7}
      & \multirow{3}{*}{Qwen-Code}
          & 1 & 0.85 & 0.94 & 0.89 & 0.05 \\
      &   & 2 & 0.85 & 0.94 & 0.89 & 0.05 \\
      &   & 3 & 0.85 & 0.94 & 0.89 & 0.05 \\
      \midrule
      \multirow{9}{*}{$x_{few-CoT}$}
      & \multirow{3}{*}{GPT-4o}
          & 1 & 0.87 & 0.83 & 0.84 & 0.12 \\
      &   & 2 & 0.87 & 0.82 & 0.84 & 0.11 \\
      &   & 3 & 0.87 & 0.82 & 0.84 & 0.12 \\
      \cmidrule(lr){2-7}
      & \multirow{3}{*}{GPT-OSS}
          & 1 & \textbf{0.91} & \textbf{0.75} & 0.82 & \textbf{0.27} \\
      &   & 2 & 0.91 & 0.75 & 0.82 & 0.26 \\
      &   & 3 & 0.91 & 0.76 & 0.83 & 0.26 \\
      \cmidrule(lr){2-7}
      & \multirow{3}{*}{Qwen-Code}
          & 1 & 0.89 & 0.49 & 0.63 & 0.11 \\
      &   & 2 & 0.89 & 0.49 & 0.63 & 0.12 \\
      &   & 3 & 0.89 & 0.52 & 0.65 & 0.12 \\
      \midrule
      
      All flaky & & & 0.85 & 1 & \textbf{0.92} & 0.00\\
      Random & & & 0.84 & 0.51 & 0.64 & 0.00\\
      \bottomrule
    \end{tabular}
    
    \label{tab:classification-idoft}
\end{table}

\begin{table}
    \caption{Flakiness classification results on the FlakeBench dataset for each prompt, separated by model. The numbers in bold represent the highest value in a column.}
    \centering
    \begin{tabular}{l l r rrrr}
      \toprule
      \textbf{Approach} & \textbf{Model} & \textbf{Iter.} & \textbf{Prec.} & \textbf{Rec.} & \textbf{F1} & \textbf{MCC} \\
      \midrule
      
      \multirow{9}{*}{$x_{zero}$}
      & \multirow{3}{*}{GPT-4o} 
          & 1 & 0.06 & 0.42 & 0.11 & 0.09 \\
      &   & 2 & 0.07 & 0.44 & 0.12 & 0.11 \\
      &   & 3 & 0.07 & 0.44 & 0.12 & 0.10 \\
      \cmidrule(lr){2-7}
      & \multirow{3}{*}{GPT-OSS}
          & 1 & \textbf{0.10} & 0.53 & 0.17 & \textbf{0.17} \\
      &   & 2 & 0.10 & 0.53 & 0.17 & 0.17 \\
      &   & 3 & 0.09 & 0.50 & 0.16 & 0.16 \\
      \cmidrule(lr){2-7}
      & \multirow{3}{*}{Qwen-Code}
          & 1 & 0.08 & 0.57 & 0.15 & 0.16 \\
      &   & 2 & 0.08 & 0.57 & 0.15 & 0.16 \\
      &   & 3 & 0.08 & 0.57 & 0.15 & 0.16 \\
      \midrule
      \multirow{9}{*}{$x_{zero-CoT}$}
      & \multirow{3}{*}{GPT-4o}
          & 1 & 0.05 & 0.7 & 0.09 & 0.09 \\
      &   & 2 & 0.05 & 0.67 & 0.09 & 0.08 \\
      &   & 3 & 0.05 & 0.70 & 0.09 & 0.08 \\
      \cmidrule(lr){2-7}
      & \multirow{3}{*}{GPT-OSS}
          & 1 & 0.06 & 0.79 & 0.11 & 0.14 \\
      &   & 2 & 0.06 & 0.79 & 0.11 & 0.14 \\
      &   & 3 & 0.06 & 0.82 & 0.11 & 0.14 \\
      \cmidrule(lr){2-7}
      & \multirow{3}{*}{Qwen-Code}
          & 1 & 0.04 & 0.95 & 0.08 & 0.09 \\
      & & 2 & 0.04 & 0.95 & 0.08 & 0.09 \\
      & & 3 & 0.04 & 0.95 & 0.08 & 0.09 \\
      \midrule
      \multirow{9}{*}{$x_{few-CoT}$}
      & \multirow{3}{*}{GPT-4o}
          & 1 & 0.04 & 0.67 & 0.07 & 0.04 \\
      &   & 2 & 0.04 & 0.70 & 0.07 & 0.05 \\
      &   & 3 & 0.04 & 0.70 & 0.08 & 0.05 \\
      \cmidrule(lr){2-7}
      & \multirow{3}{*}{GPT-OSS}
          & 1 & 0.06 & 0.88 & 0.11 & 0.15 \\
      &   & 2 & 0.06 & 0.88 & 0.11 & 0.15 \\
      &   & 3 & 0.06 & 0.88 & 0.11 & 0.15 \\
      \cmidrule(lr){2-7}
      & \multirow{3}{*}{Qwen-Code}
      & 1 & 0.09 & \textbf{1} & \textbf{0.17} & 0.06 \\
      & & 2 & 0.10 & 1 & 0.18 & 0.06 \\
      & & 3 & 0.09 & 1 & 0.17 & 0.06 \\
      \midrule
      All flaky & & & 0.03 & 1 & 0.06 & 0.00\\
      Random & & & 0.03 & 0.49 & 0.05 & 0.00\\
      \bottomrule
    \end{tabular}
    \label{tab:classification-flakebench}
\end{table}

\section{Results}
\subsection{RQ1: Classification}
\label{sec:results-classification}
We report the results of the classification across three iterations in \Cref{tab:classification-idoft} for \emph{IDoFT} and \Cref{tab:classification-flakebench} for \emph{FlakeBench}.

On the \emph{IDoFT} dataset, $x_{few-CoT}$ achieved the best result for all models. While Qwen-Code achieved similar results for $x_{zero}$ and $x_{zero-CoT}$, the two general-purpose models achieved higher results with $x_{few-CoT}$ by a large margin. These results suggest that both general-purpose models picked up relevant patterns from the examples in the context, thereby improving their overall performance. GPT-4o achieved an MCC of 0.12 with $x_{few-CoT}$, representing an improvement of a factor of 4 compared to other prompting techniques. GPT-OSS improved on the results of GPT-4o for all prompting techniques, achieving an MCC of 0.27 for $x_{few-CoT}$. However, an MCC of 0.27 still only indicates a weak correlation between the ground-truth labels and the model's predictions.

For the \emph{FlakeBench} dataset, GPT-OSS also yielded the best results with every prompt. Comparing prompting techniques, $x_{zero}$ achieved the best result on FlakeBench with an $MCC$ of 0.17, in contrast to IDoFT. However, while $x_{few-CoT}$ improved the results of $x_{zero}$ by a large margin on IDoFT, the two prompts resulted in similar results on the FlakeBench dataset ($MCC=0.17$ vs. $MCC=0.15$). $x_{zero-CoT}$ achieved the worst results on both datasets. Similar to the other two prompts, however, the differences in performance are less pronounced on the FlakeBench dataset. On FlakeBench, we observed a 0.03 decrease in performance for $x_{zero-CoT}$. On the IDoFT dataset, the largest margin between prompts is -0.30 ($x_{zero-CoT}$ vs. $x_{few-CoT}$).

We observed only minor differences in the performance metrics between different experiment runs. In most cases, there was no difference greater than 0.01 in $MCC$ between the three repetitions. The largest range was between $-0.03$ and $0.01$ for $x_{zero-CoT}$ on IDoFT.

\begin{tcolorbox}[enhanced jigsaw,sharp corners, drop fuzzy shadow=ShadowColor]
\textbf{Answer to RQ1}: We found that LLMs show a poor performance in classifying tests as flaky (or not) on both datasets. On \emph{IDoFT}, we obtained the best result with few-shot prompting, achieving an $MCC$ of 0.27. By F1 score, none of the approaches surpassed the \emph{All flaky} baseline on IDoFT. On \emph{FlakeBench}, the zero-shot prompt achieved the best result with an $MCC$ of $0.17$.
\end{tcolorbox}

\subsection{RQ2: Reliability}
We report the normalized Hamming distances between three repetitions in \Cref{tab:hamming-flakebench} for \emph{FlakeBench} and \Cref{tab:hamming-idoft} for \emph{IDoFT}. 

As shown in \Cref{tab:hamming-flakebench} and \Cref{tab:hamming-idoft}, $x_{zero-CoT}$ exhibited the highest non-determinism for GPT-4o and GPT-OSS on both datasets, resulting in normalized Hamming distances of up to 0.25. For GPT-4o, the difference between $x_{zero-CoT}$ and the results for the other two prompts was more pronounced, with $x_{zero-CoT}$ showing an increase in the normalized Hamming distance by at least a factor of 2. For GPT-OSS, the non-determinism increased by a factor of 1.07 for $x_{zero-CoT}$. 

For Qwen-Code, $x_{few-CoT}$ resulted in the highest level of non-determinism on both datasets. Compared to the two GPT models, Qwen-Code exhibited a lower level of non-determinism overall. Furthermore, Qwen-Code was the only model that achieved equal results between two repetitions for $x_{zero}$ and $x_{zero-CoT}$ on the FlakeBench dataset. 

\begin{tcolorbox}[enhanced jigsaw,sharp corners, drop fuzzy shadow=ShadowColor]
\textbf{Answer to RQ2}: We found that non-determinism affects LLM-based flakiness classification. The non-determinism in our experiments ranged from 0\% to 25\% normalized Hamming distance. There were differences in non-determinism among models, datasets, and prompting techniques. Qwen-Code exhibited the lowest degree of non-determinism overall. GPT-OSS showed more non-determinism than GPT-4o. 
\end{tcolorbox}

\subsection{RQ3: Manual Investigation}

\begin{figure}
  \centering
  \includegraphics[width=\columnwidth]{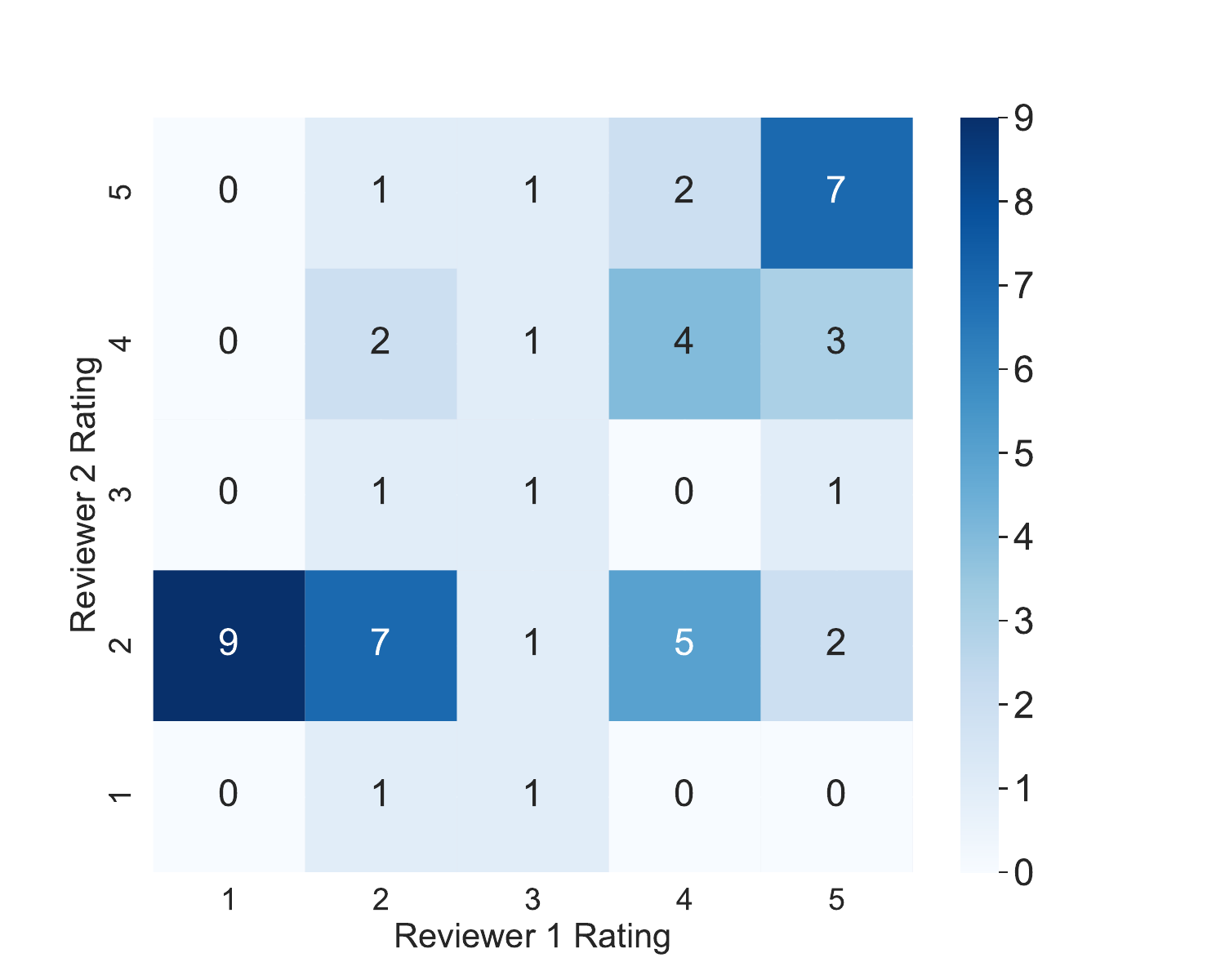}
  \caption{Heatmap showing the results of the survey from two reviewers. 
  Answers to the question \enquote{How likely would a developer proficient in Java classify this test as flaky?} range from 1 (Very likely) to 5 (Very unlikely).}
  \label{fig:heatmap-reviewers} 
\end{figure}

\Cref{fig:heatmap-reviewers} shows the labeling results from the two reviewers.
Overall, the reviewers achieved a Kappa score of 0.57, indicating moderate agreement. Both reviewers produced an average rating close to the center of the scale (Reviewer 1: 3.08, Reviewer 2: 3.14). 
Rating 3 was used least frequently in only 8 out of 100 ratings, i.e., reviewers tended to make more decisive judgments towards the two ends of the scale.
The most common rating was 2~(Likely) for 36 out of 100 ratings, followed by 5~(Very unlikely), which was used in 24 out of 100 ratings.
As shown in \Cref{tab:labels-fix}, the reviewer ratings varied across the fix categories described in \Cref{tab:fix-labels}. Reviewers considered changes to assertions or data structures in the test code easier to detect with an average rating of 2 (Likely). In contrast, the categories \texttt{reset\_variable} and \texttt{reorder\_parameters} received a rating of 4 (Unlikely). 

\Cref{fig:examples} illustrates two examples from the reset\_variable and change\_assertion. In \Cref{sub:a}, the test was fixed by resetting the \texttt{iterations} attribute of the \texttt{Enums} class. However, the existence of this attribute was not apparent from the test code. Furthermore, the increase of the variable happens as a side-effect of the call of \texttt{defaultPropertyTrialCount}, which was not defined in the test code either. Therefore, reviewers considered it unlikely that the test could be classified as flaky solely based on an understanding of the test code and without additional context. In contrast, in \Cref{sub:b}, the flakiness was caused by a mistakenly assumed order on the result of a stringified JSON representation. Since stringifying JSON does not guarantee an ordered representation of the key-value pairs, the test code was sufficient to identify and fix the issue. Both reviewers agreed that cases similar to these two example cases were common in the annotated test set. This suggests that some flakiness issues may be easier to detect, while others require additional context for clearance.

\begin{tcolorbox}[enhanced jigsaw,sharp corners, drop fuzzy shadow=ShadowColor]
\textbf{Answer to RQ3}: Whether humans consider themselves capable of identifying a test as flaky based on the test code depends on the type of flakiness. While some types of flakiness were apparent in the test code, e.g., the reliance on the order of an unordered collection in the test, other types, such as problems caused by side effects of the tested functionality or test utility functions, were not visible in the test code. 
\end{tcolorbox}

\begin{table}
\renewcommand{\arraystretch}{0.9}
  \caption{Normalized Hamming distances between iterations of the same prompt for the FlakeBench dataset.}
  \centering
  \setlength{\tabcolsep}{4pt}
  \begin{tabular}{l l l l l}
    \toprule
     \textbf{Model} & \textbf{Prompt} & $H(y_1, y_2)$ & $H(y_2, y_3)$ & $H(y_1, y_3)$ \\
    \midrule
    \multirow{3}{*}{GPT-4o}  
      & $x_{zero}$     & 0.08 \phantom{-}(720)        & 0.08 \phantom{-}(667)        & 0.08 \phantom{-}(667) \\
      & $x_{zero-CoT}$ & \textbf{0.16 (\num{1401})} & \textbf{0.25 (\num{2134})} & \textbf{0.25 (\num{2307})} \\
      & $x_{few-CoT}$    & 0.06 \phantom{-}(518)        & 0.06 \phantom{-}(477)        & 0.07 \phantom{-}(605) \\
    \midrule
    \multirow{3}{*}{GPT-OSS} 
      & $x_{zero}$     & 0.06 \phantom{-}(502)        & 0.07 \phantom{-}(596)        & 0.06 \phantom{-}(556) \\
      & $x_{zero-CoT}$ & \textbf{0.13 (\num{1130})} & \textbf{0.13 (1192)} & \textbf{0.13 (\num{1150})} \\
      & $x_{few-CoT}$    & 0.12 \phantom{-}(\num{992})        & 0.12 \phantom{-}(\num{999})        & 0.11 \phantom{-}(\num{971}) \\
      \midrule
    \multirow{3}{*}{\shortstack[l]{Qwen-\\Code}} 
      & $x_{zero}$     & 0.01 \phantom{--}(120)        & 0.00 \phantom{----}(0)        & 0.01 \phantom{--}(120) \\
      & $x_{zero-CoT}$ & 0.00 \phantom{-----}(5) & 0.00 \phantom{----}(0) & 0.00 \phantom{-----}(5) \\
      & $x_{few-CoT}$    & \textbf{0.13 (\num{1104})}        & 0.00 \phantom{----}(\textbf{\num{1}})        & \textbf{0.13 (\num{1103})} \\
    \bottomrule
  \end{tabular}
  \label{tab:hamming-flakebench}
\end{table}

\begin{table}
  \caption{Normalized Hamming distances between iterations of the same prompt for the IDoFT dataset.}
  \centering
  
  \begin{tabular}{l l l l l}
    \toprule
     \textbf{Model} & \textbf{Prompt} & $H(y_1, y_2)$ & $H(y_2, y_3)$ & $H(y_1, y_3)$ \\
    \midrule
    \multirow{3}{*}{GPT-4o}  
      & $x_{zero}$     & 0.01 \phantom{-}(52)        & 0.01 \phantom{-}(48)        & 0.01 \phantom{-}(52) \\
      & $x_{zero-CoT}$ & \textbf{0.09 (344)} & \textbf{0.09 (362)} & \textbf{0.09 (366)} \\
      & $x_{few-CoT}$    & 0.02 \phantom{-}(96)        & 0.02 \phantom{-}(83)        & 0.02 \phantom{-}(87) \\
    \midrule
    \multirow{3}{*}{GPT-OSS} 
      & $x_{zero}$     & 0.12 (473)        & 0.12 (463)        & 0.12 (464) \\
      & $x_{zero-CoT}$ & \textbf{0.15 (583)} & \textbf{0.15 (580)} & \textbf{0.15 (575)} \\
      & $x_{few-CoT}$    & 0.13 (496)        & 0.13 (524) & 0.12 (472) \\
    \midrule
    \multirow{3}{*}{\shortstack[l]{Qwen-\\Code}}
      & $x_{zero}$     & 0.01 \phantom{-}(21)        & 0.00 \phantom{-}(19)        & 0.01 \phantom{-}(20) \\
      & $x_{zero-CoT}$ & 0.00 \phantom{--}(2) & 0.00 \phantom{--}(2) & 0.00 \phantom{--}(2) \\
      & $x_{few-CoT}$    & \textbf{0.01 \phantom{-}(36)} & \textbf{0.01 \phantom{-}(36)} & \textbf{0.01 \phantom{-}(36)} \\
    \bottomrule
  \end{tabular}
  \label{tab:hamming-idoft}
\end{table}

\begin{table}
    \caption{Average ratings aggregated by fix category.}
    \centering
    \begin{tabular}{l r r}
        \toprule
         Fix category & Support & Mean \\
         \midrule
         change\_assertion & 14 & 2.2\\
         change\_condition & 8 & 2.9\\
         misc & 6 & 4.4\\
         change\_data\_structure & 6 & 2.7\\
         reset\_variable & 5 & 3.9\\
         reorder\_parameters & 3 & 4.2\\
         reorder\_data & 3 & 3.0\\
         change\_assertion,change\_condition & 2 & 3.5\\
         reset\_variable,handle\_exceptions & 2 & 4.0\\
         reorder\_data,change\_data\_structure & 1 & 3.0\\
         reorder\_data,change\_assertion,change\_condition & 1 & 2.0\\
         sleep & 1 & 5.0\\
         change\_data\_format & 1 & 1.5\\
         \bottomrule
    \end{tabular}
    \label{tab:labels-fix}
\end{table}

\begin{figure*}
\begin{subfigure}{\textwidth}
\begin{lstlisting}[language=javadiff, 
  basicstyle=\small\ttfamily,
  captionpos=b, 
  label={lst:unordered-collection-example}]
public void enums() throws Exception {
  assertThat(testResult(Enums.class),isSuccessful());
  assertEquals(defaultPropertyTrialCount(),Enums.iterations);
  assertEquals(EnumSet.of(HALF_UP,HALF_EVEN),new HashSet<>(Enums.values.subList(0,2)));
+ Enums.iterations=0;
}
\end{lstlisting}
\caption{Side-effects can modify \texttt{Enums.iterations}, which is not apparent in the test. Reviewers considered a flaky classification unlikely.}
\label{sub:a}
\end{subfigure}

\vspace{\baselineskip}

\begin{subfigure}{\textwidth}
\begin{lstlisting}[language=javadiff]
public void TestBlikDetailsSerialization() throws JsonProcessingException {
  String expectedJson=""{...}"";
  BlikDetails blikDetails=new BlikDetails();
  blikDetails.setBlikCode(""blikCode"");
  PaymentsRequest paymentsRequest=createPaymentsCheckoutRequest();
  paymentsRequest.setPaymentMethod(blikDetails);
  String gson=GSON.toJson(paymentsRequest);
  assertEquals(expectedJson,gson);
  String jackson=OBJECT_MAPPER.writeValueAsString(paymentsRequest);
- assertEquals(expectedJson,jackson);
+ assertJsonStringEquals(expectedJson,jackson);
}
\end{lstlisting}
\caption{JSON does not guarantee an order when stringified. Reviewers agreed that it is likely to detect the test as flaky.}
\label{sub:b}
\end{subfigure}
\caption{Examples of \texttt{reset\_variable} and \texttt{change\_assertion} (diff indicates fixed version of the corresponding test).}
\label{fig:examples}
\end{figure*}

\section{Discussion}
\label{sec:discussion}
In this section, we discuss our empirical results.

\textbf{On the flakiness classification.}
We found that LLMs without fine-tuning struggle to classify tests as flaky solely based on the test code. 
Achieving an $MCC$ of 0.27 in the best case, we argue that the performance is far below the requirements for deploying such a model in a production scenario. 
This is further aggravated by the exhibited non-determinism of the models when asked to classify a test. 
With the aim of deploying a flakiness detection model in practice in mind, it is crucial that the model reliably detects flaky tests. 
Based on their experience with the deployment of other AI systems in development processes, practitioners at SAP noted that developers quickly lose trust in a system that delivers inconsistent results for equal inputs~\cite{baltes2025rethinking}. We envision two possible future avenues to solve these problems. 
Firstly, to improve the models' predictive performance, future work could experiment with additional context, such as the code under test, documentation, or test helper functions. 
Based on preliminary manual experiments with examples from the IDoFT dataset, we found that presenting the model with the \emph{relevant} context, which contains all necessary information, can improve the model's predictions.
We conjecture that agentic systems or retrieval-augmented generation (RAG) approaches that dynamically fetch relevant information may yield better results than those observed in our study.
Secondly, to mitigate the negative effects of the model's non-determinism, future work could experiment with approaches such as self-consistency, ensembles, or combinations of the two~\cite{wang2023selfconsistency}. 

\textbf{On the labeling results.} 
Based on our manual labeling of flaky tests in the IDoFT dataset, we found that classifying tests as flaky only based on the given test code may be infeasible for certain types of flakiness.
Given the fix labels provided by previous work~\cite{fatima2024flakyfix}, we observed differences in the feasibility of classifying a test as flaky for tests with different fix labels.
More specifically, while it may be possible to classify a test as flaky based on the existence of trivial test smells, such as the use of unordered collections or sleep statements in the test code, identifying the absence of proper setup or teardown requires further information about the code being tested.

This finding is also in line with previous research on fixing flaky tests~\cite{chen2024flakiness,chen2024neurosymbolic}. 
When using GPT-4 to fix flaky tests only given the test code, Chen et al.~\cite{chen2024flakiness} found that performance varies between different types of flakiness.
Although their approach successfully repaired 58\% of flaky tests, whose flakiness was caused by implementation errors in the test code, the authors pointed out that GPT-4 was not capable of repairing other types of flakiness in their study. 

Further research is required to identify the differences between test-only flakiness issues and more complex issues that occur in other parts of the source code. Based on our analysis of the IDoFT dataset, we conjecture that these different types of flakiness may also have varying impacts on testing practices. Although test-only issues can be easily identified and fixed by developers, more sophisticated issues can be challenging to debug and fix, as they often rely on more complex program states~\cite{lam2020understanding}.

\section{Threats to Validity}
\label{sec:threats}
\subsection{Internal Validity}
In this section, we describe possible alternative explanations for our results~\cite{brewer2000research}.

\textbf{On the LLM results.} We examined a limited set of prompting techniques and LLMs. Other models or prompts might yield different results. Furthermore, the non-determinism of LLMs could have affected the results of our study.
To mitigate this threat, we used greedy decoding and repeated our experiments three times~\cite{baltes2025guidelinesempiricalstudiessoftware}. Furthermore, we added more depth to the study by manually investigating 50 samples of the dataset. The results of our manual investigation align with the findings of our experiments using LLMs. 

\textbf{On the manual investigation.} We manually examined 50 samples of the IDoFT dataset. We randomly sampled these 50 samples. Thus, the resulting sample may have suffered from sampling bias, which might lead to an overestimation of the proportion of tests in which flakiness is not apparent in the test code.

\subsection{External Validity}
We describe the extent to which our results generalize to other contexts~\cite{baltes2022sampling}.

\textbf{On the employed datasets.} Since we only used two datasets with a limited number of flaky tests written only in Java, our results may not generalize to all types of tests. However, we note that the results align with our experience in real-world CI/CD pipelines. For example, the ticket system at SAP offers a dedicated tag to flag test-only issues, where the fault lies within the test code. However, not all flakiness-related issues are tagged as test-only. In fact, differentiating between test-only issues and more complex flakiness issues is a common challenge.

\subsection{Construct Validity}
We describe the degree to which our metrics measure the intended properties~\cite{ralph2018construct}.

\textbf{On the evaluation metrics for the flakiness classification.} Similar to prior studies on flakiness classification, we report the precision, recall, and F1 score for our prediction results~\cite{fatima2022flakify,more2025analysis,rahman2024quantizing}. Since the class distributions of the two datasets in this study are opposed, this results in measures that are hardly comparable. However, to mitigate this problem, we added the MCC, a common metric in the context of binary classification tasks, to enable a comparison of the results on the two datasets~\cite{chicco2020advantages}. 

\section{Conclusions}
\label{sec:conclusions}
We evaluated three LLMs without additional fine-tuning for flakiness classification, given only the test code.
Based on evaluations across two benchmark datasets, our results show that LLMs struggle to correctly classify flaky tests.
Our manual analysis of 50 samples suggests that this is due to the absence of necessary information in the test code for correct classification.
This finding contradicts previously reported results from fine-tuning approaches, which suggested that LLM-based flakiness classification based solely on test code is promising.

While related work has already identified issues in a fine-tuning study, our examination of the replication package for another related study suggests that the relatively high reported scores are due to two reasons. 
On the one hand, a skewed evaluation setup in which simple baseline approaches already yield high scores. Second, we expect the high scores in the reported metrics to be due to overfitting to overly simple characteristics that differentiate flaky tests from non-flaky tests in the given dataset.

Furthermore, we evaluated the non-determinism of LLMs for flakiness classification. Our results indicate that the non-determinism varies between models, prompts, and datasets. 
We argue that the exhibited amount of non-determinism is problematic for deploying such models in practice.

Our findings can motivate future research evaluating general-purpose LLMs with additional sources of information beyond the test code. For this, we mainly see potential in the use of agentic systems or RAG, inspired by previous work on program repair~\cite{bouzenia2024repairagent,rondon2025evaluating}.



\balance
\bibliographystyle{IEEEtran}
\bibliography{references}

\end{document}